\documentclass[letterpaper, 10 pt, conference]{ieeeconf}  

\IEEEoverridecommandlockouts                              

\usepackage{graphics} 
\usepackage{amsmath,amssymb,amsfonts}
\usepackage{algorithmic}
\usepackage{graphicx}
\usepackage{textcomp}
\usepackage{xcolor}
\usepackage{biblatex}
\title{\LARGE \bf
The Lost-K and Shorter-J Phenomenon in Non-Standard Ballistocardiography Data*
}

\author{Shuai Jiao$^{1}$, Jian Fang$^{1}$, Tianshu Zhou$^{1}$, Jinsong Li$^{1}$,Yanhong Liu$^{2}$, Ye Liu$^{3}$,Ming Ju$^{4}$
\thanks{This work was surpported by Zhejiang Lab and Beijing LovearthTech Co.,Ltd.}
\thanks{$^{1}$Shuai Jiao, Jian Fang, Tianshu Zhou, Jinsong Li are with Zhejiang Lab, Hangzhou, China.
Email:jiaoshuai@zhejianglab.org,
fangjian@zhejianglab.org,
zhouts@zhejianglab.org,
ljs@zhejianglab.org}%
\thanks{$^{2}$Yanhong Liu is with the Department of Anesthesiology, The First Medical Center of Chinese PLA Hospital. Beijing China. Email: 18618338301@163.com}%
\thanks{$^{3}$Ye Liu is with the Heart Center, Beijing Chaoyang Hospital. Beijing China. Email: 19970059@ccmu.edu.cn}%
\thanks{$^{4}$Ming Ju is with the Heart Center, Medical Device R\&D Department, Beijing LovearthTech Co.,Ltd. Beijing China. Email: juminghit@gmail.com}%
}

\begin{document}

\maketitle
\thispagestyle{empty}
\pagestyle{empty}

\begin{abstract}
Non-standard ballistocardiogram(BCG) data generally do not have prominent J peaks. This paper introduces two phenomena that reduce the prominence of J peaks: the shorter-J phenomenon and the lost-K phenomenon, both of which are commonly observed in non-standard BCG signals . This paper also proposes three signal transformation methods that effectively improve the lost-K and shorter-J phenomena. The methods were evaluated on a time-aligned  ECG-BCG dataset with 40 subjects. The results show that based on the transformed signal, simple J-peak-based methods using only the detection of local maxima or minima show better performance in locating J-peaks and extracting BCG cycles, especially for non-standard BCG data.
\newline

\end{abstract}

\section{INTRODUCTION}

Ballistocardiogram (BCG) is a non-contact and non-invasive method that is used for recording the body’s recoil response to the forces created by the cardiovascular blood pumping activity\cite{starr1946further,baker1950coronary}.Attributed to the mechanism of BCG, physiological information of the human cardiovascular system can be obtained in a non-invasive way, which can significantly reduce the psycho-physiological discomfort of patients.

The BCG  and its cycle waveform are useful for learning about cardiac cycle physiology. Despite major disparities, most BCG setups produce similar signal shapes. An example of time-aligned ECG-BCG signal and its components for three cardiac cycles are shown in Figure\ref{fig_bcg_cycle}. The “I”, “J”, and “K” waves are typical of BCG recordings and have been proved with clinical signiﬁcance in cardiovascular parameters \cite{kim2016ballistocardiogram}. The first major negative deflection after the ECG QRS complex is called the I wave and is followed by the most dominant peak, the J wave. The J-wave is the largest headword wave that occurs late in systole. The most prominent peak in each J-wave is designated  as the J-peak, which consistently represents the heartbeat as the R-peak in an electrocardiogram (ECG). Consequently, the precise location of the J-peak is of fundamental importance for the estimation of inter-beat intervals (IBI), heart rate (HR) and the acquisition of other physiological information.

\section{ MOTIVATION}

\subsection{Dataset}
\begin{figure*}[htbp]
\centerline{\includegraphics[width=\linewidth]{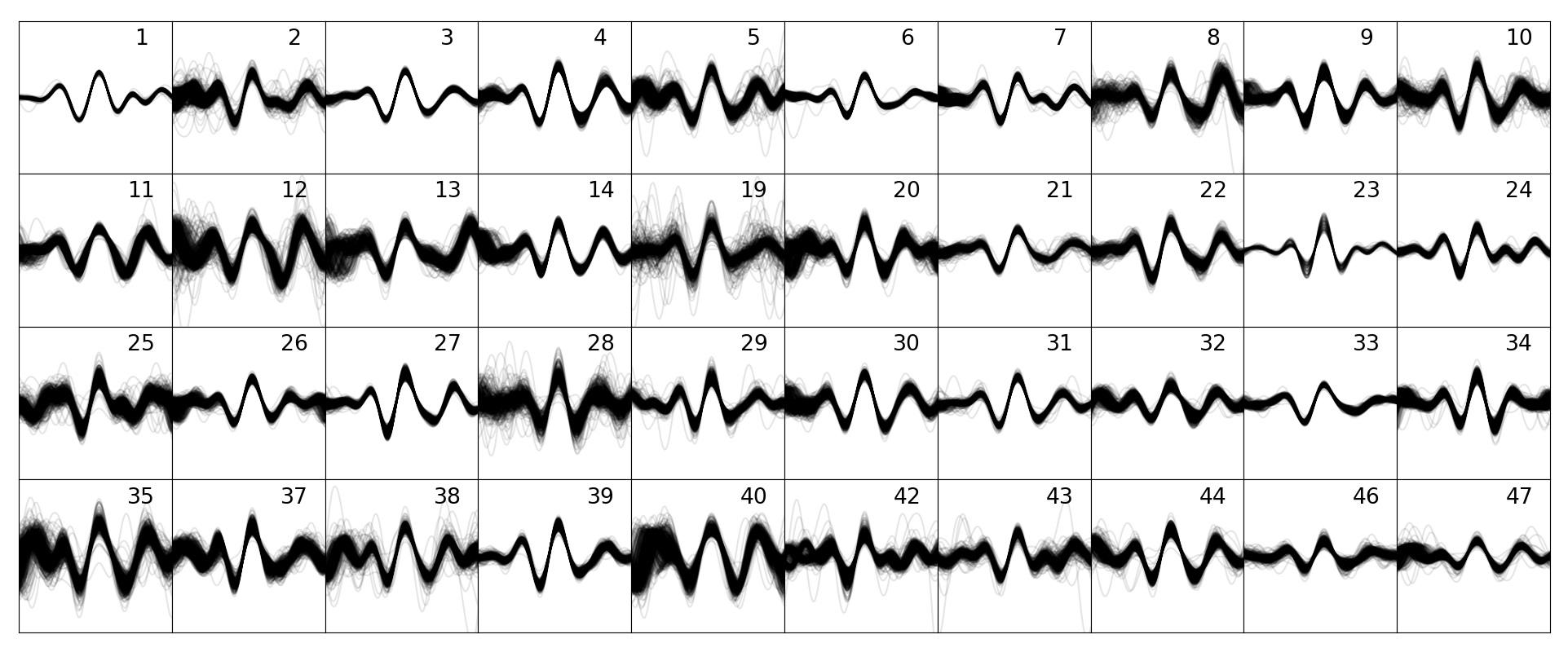}}
\caption{Stacked view of BCG waveform cycles for the refereed 40 records of the datasets. Signals are bandpass filtered using 2-6Hz frequency range before segmenting manually with the help of time-aligned ECG signal. }
\label{fig:waves_stack}
\end{figure*}

This paper utilises a publicly available dataset \cite{Carlson2021Bed-BasedBallistocardiography} acquired from a multi-channel bed-based BCG system.The dataset encompasses data from 40 participants of varying ages and body types. The rationale for this choice is based on the following considerations:
\begin{itemize}
\item To the best of the authors’ knowledge, no other publicly available dataset includes time-aligned ECG, PPG, BCG, and continuous blood pressure data.
\item The timing of the ECG and BCG are strictly synchronised and the phase-delay being evaluated as accurate.
\item The dataset includes additional time-aligned ground-truth cardiopulmonary data (DP, SP and SV), which could be utilised for further research. This could include bed-BCG-based blood pressure tracking or stroke volume estimation. 
\end{itemize}

We selected the Film0 data from each participant as it exhibited the optimal signal quality. In the following sections of this paper, X1001 denotes the data from the X1001 participant and sensor Film0, for the sake of brevity.  Figure \ref{fig:waves_stack} shows a stacked plot of all waveform cycles for the 40 records in the dataset.

\subsection{lost-K phenomenon}

\begin{figure}[htbp]
\centerline{\includegraphics[width=\linewidth]{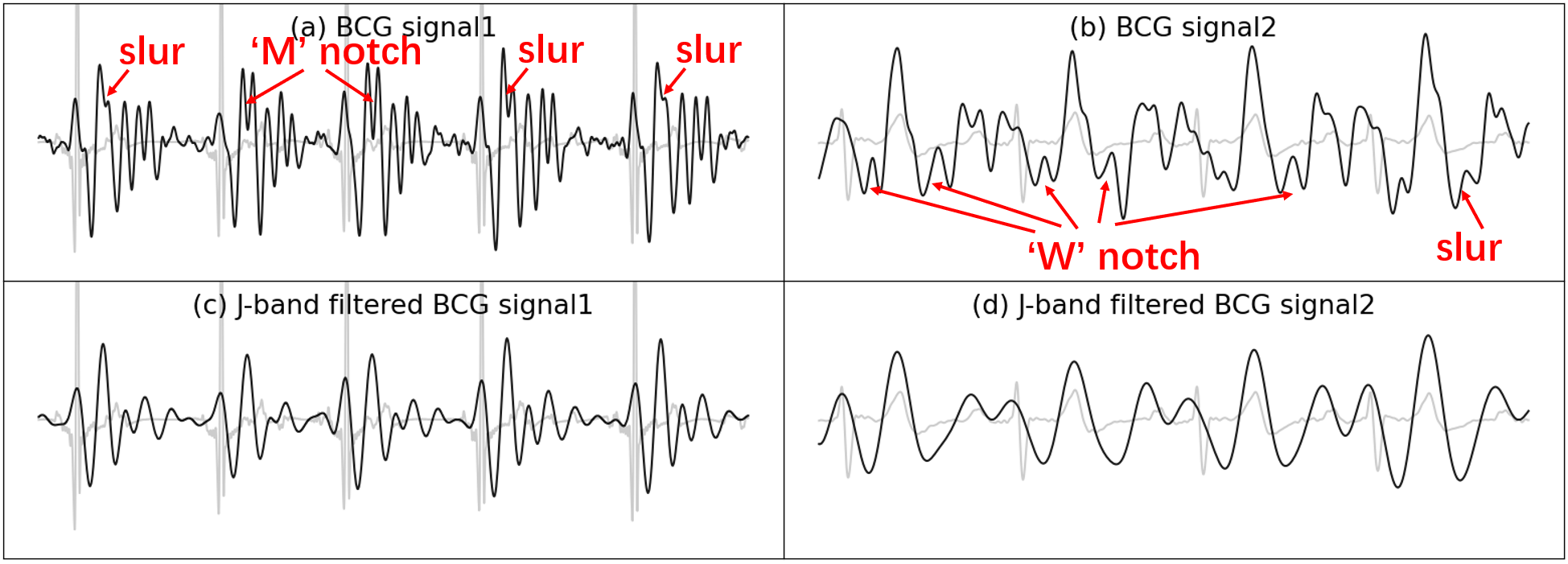}}
\caption{Fragments of signal from X1001(a) and X1005(b) before and after j-band filter. Before filtering, slurs and notches happen around J peak location(a) and I and K valley(b), making them difficult to identify. (c) and (d) are the corresponding j-band filtered signals, where the i, j, and k waveforms are readily discernible. }
\label{fig_slur_notch}
\end{figure}

Slurs and notches are frequently observed in non-standard BCG signals, particularly the 'M' notch and 'W' notch, which distort the I, J, and K waveform pattern, rendering them challenging to identify. Lixin (2016) \cite{Lixin2016JpeakExtraction} found that the majority of J-peak energy resides within a narrow frequency band, designated as the J-peak band. Subsequently, the author proposed the implementation of a J-band passing filter (1-6 Hz), which significantly reduces the slur and notches, facilitating the identification of the I-J-K waveforms.Fig \ref{fig_slur_notch} the benefit of j-band filter. 

\begin{figure}[htbp]
\centerline{\includegraphics[width=\linewidth]{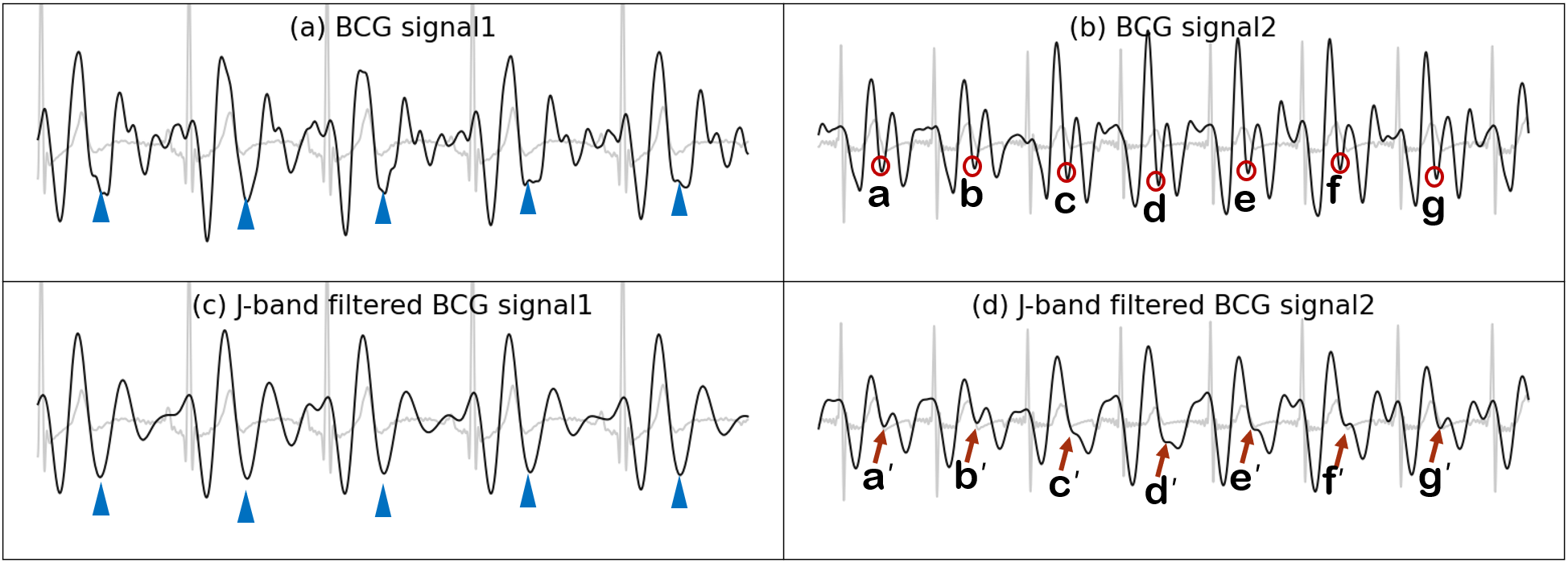}}
\caption{ A fragment from X1003 (a) and X1021(b) and the corresponding j-band filtered signal (c) and (d). The j-band filter for X1003 shows negligible shape loss, as depicted by the $\Delta$ symbols. In contrast, the J-band filter for X1021 results in significant shape loss, with some k-valleys becoming less noticeable (`a-a'`, `b-b', and `g-g'`), while others disappear entirely, leaving as notches and slurs (`c-c'`, `d-d'`, `e-e'` and `f-f'`). }
\label{fig_lostk}
\end{figure}

However, for j-band filter, the abandonment of signal energy in the 6-10 Hz range results in the loss of noticeable K-valley waveforms in certain instances. We called the losing K-valley phenomenon as the "lost-K" phenomenon.  Figure \ref{fig_lostk} illustrates the lost-K phenomenon in two different dataset recordings. The j-band filter is effective for X1003, yielding clean I-J-K waveforms. In contrast, it works badly for X1021, resulting in the loss of most notable K-valleys.

The lost-K phenomenon is not a significant issue for BCG segmentation and cycle extraction, as the majority of algorithms rely on the prominent J peak or the I-J waveform. However, some j-peak location methods, as outlined in reference \cite{Samuel2019Ejection}, which rely on the K valley, will not be applicable in this context. Besides, the lost-K phenomenon is detrimental to the majority of research projects that rely on K-valley location for BCG-based pressure tracking \cite{Kim2018CufflessBloodPressure,Yousefian2020PulseTransitTime,SuBoyu2019MonitoringtheRelativeBlood}.

\subsection{shorter-J phenomenon}

\begin{figure}[htbp]
\centerline{\includegraphics[width=\linewidth]{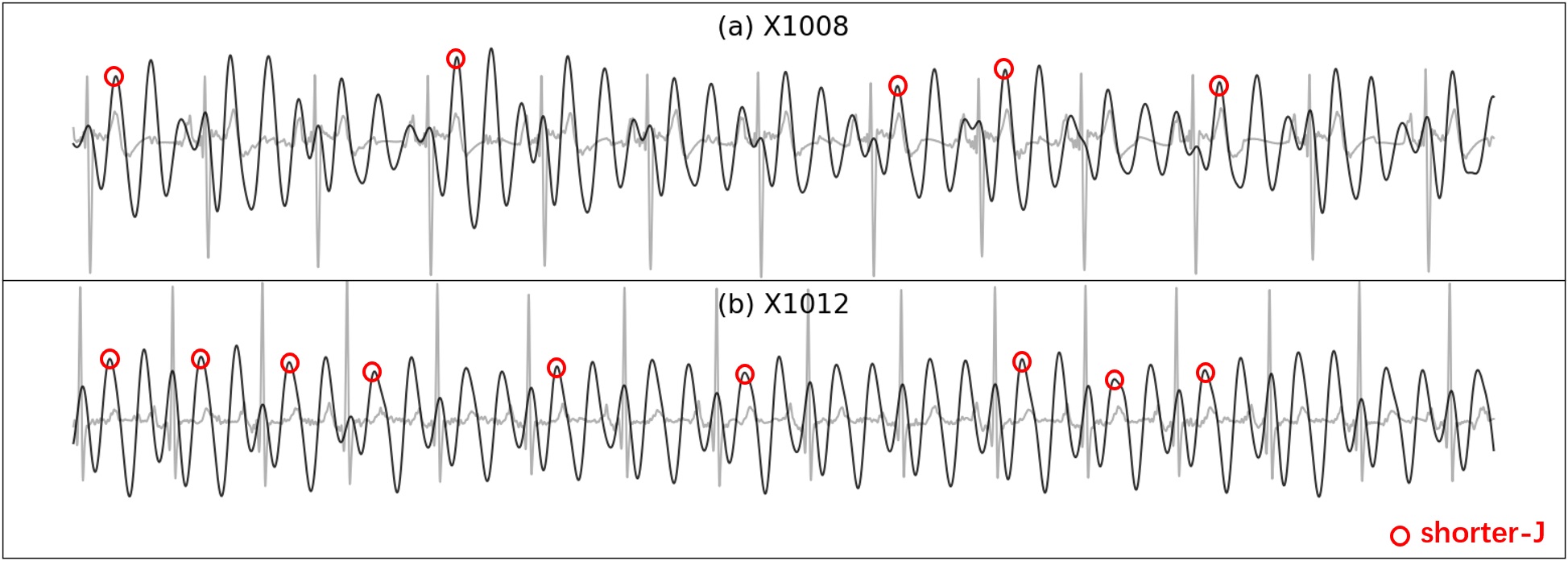}}
\caption{shorter-j phenomenon observed in X1008(a), X1011(b) indicated by the red circles. The gray line is the time-aligned ECG signal. }
\label{fig_shorterj}
\end{figure}

This section introduces the shorter-J phenomenon. In a typical H-I-J-K-L BCG waveform complex, if the J-peak is shorter in amplitude than the H-peak (previous maximum) or L-peak (subsequent maximum), the J waveform is classified as shorter-J. Figure \ref{fig_shorterj} illustrates shorter-J waveforms in 2 recordings. 

\begin{figure}[htbp]
\centerline{\includegraphics[width=\linewidth]{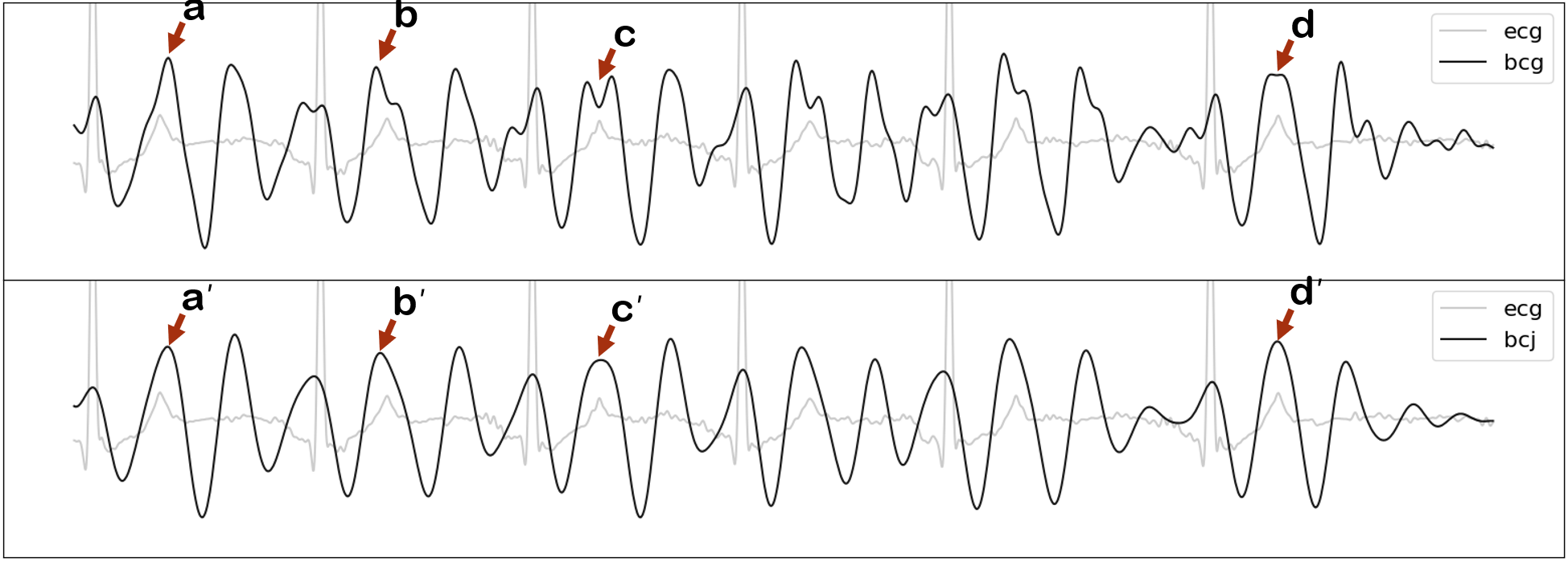}}
\caption{ Comparison of A fragment of BCG data from X1011. top is the BCG data and bottom is the j-band filtered data. as indicated, by j-band filtering, non-shorter-j in bcg are turned into shorter-j (`a-a'`,`b-b'`) in bcj; shorter-j in bcg is preserved in bcj(`c-c'`); shorter-j in bcg is improved into standard j in bcj(`d-d'`) }
\label{fig_shorterj_source}
\end{figure}

Shorter-J phenomenon is not  introduced by the j-band filter alone. In fact, some shorter-j exist in the original BCG data. however, the j-band filter distorts the BCG cycle shape, and aggregates the short-j issue. It introduces new shorter-j waveforms, which makes the BCG data more "non-standard". Figure \ref{fig_shorterj_source} shows three different effects of j-band filter, some shorter-j are new by the filtering ,some shorter-j are preserved, some are repaired.

\begin{figure}[htbp]
\centerline{\includegraphics[width=\linewidth]{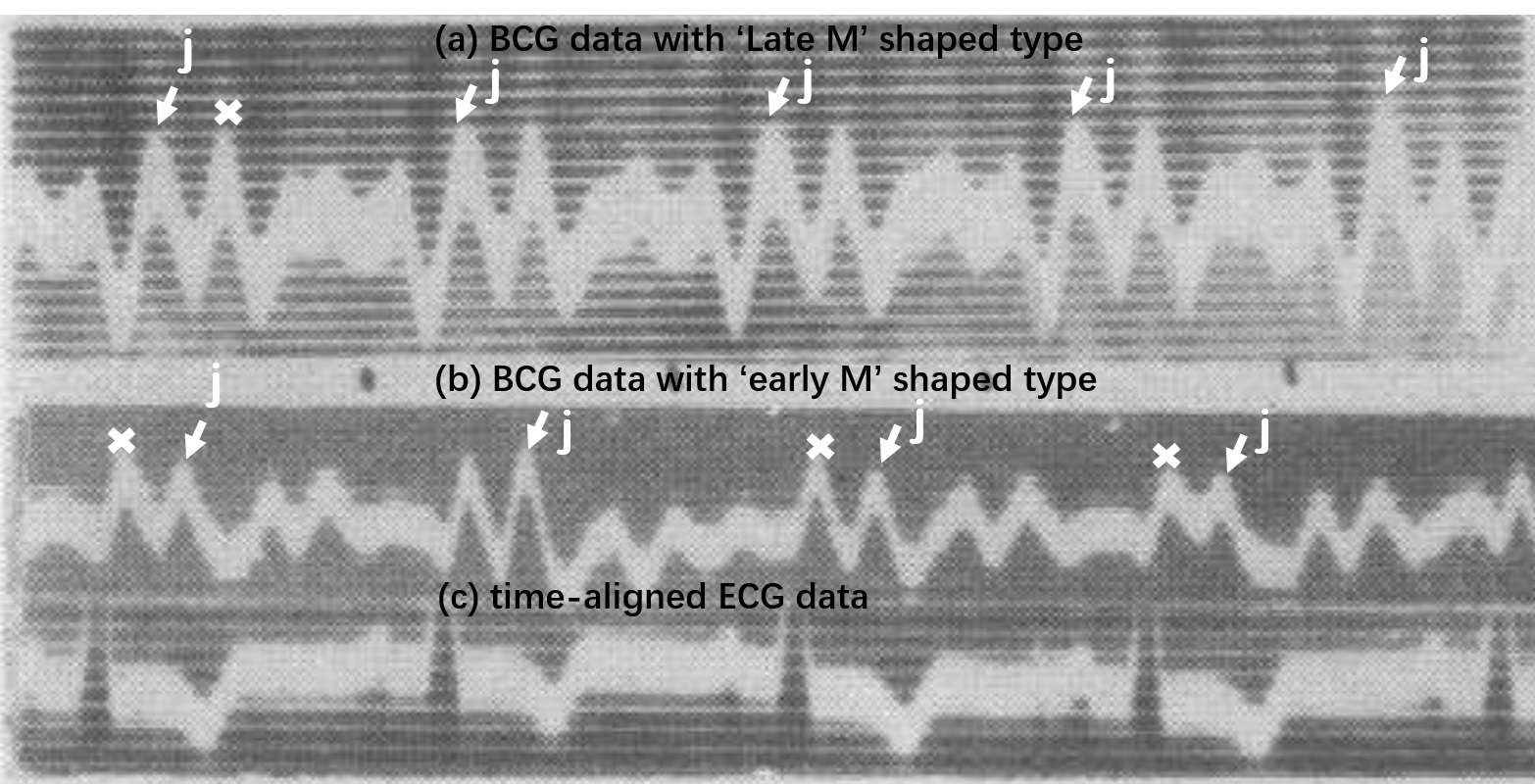}}
\caption{ (a)BCG data with early-M abnormal shaped waveform, (b)BCG data with late-M abnormal shaped waveform.(c) the corresponding ECG data. The symbol `×` denotes the shorter-J phenomenon, whereas arrows and the subsequent `j` indicate the j-peak location. Image adapted from \cite{starr1940Ballistocardiogram}.}
\label{fig_starr_abnormal}
\end{figure}

The mechanism underlying the shorter-J phenomenon remains unknown. however, it is more readily discernible in the abnormal forms observed by Starr in \cite{starr1940Ballistocardiogram}. Figure \ref{fig_starr_abnormal} illustrates the two abnormal shape types initially  described by Starr. It is evident that shorter-J is more prevalent in early-M and late-M waveforms. 

\begin{figure}[htbp]
\centerline{\includegraphics[width=\linewidth]{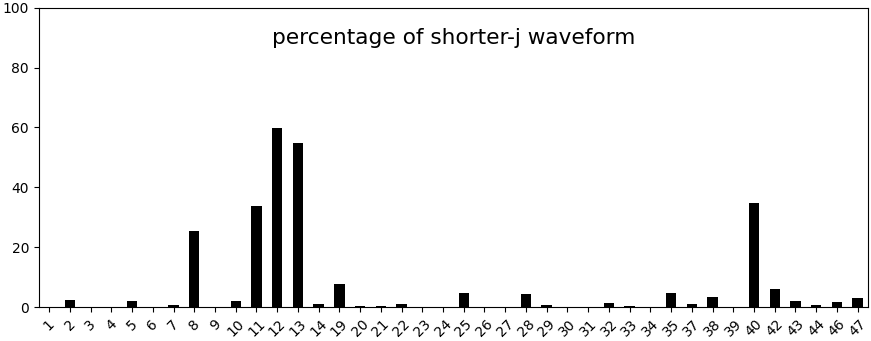}}
\caption{ shorter-j percentage across 40 dataset records. }
\label{fig_shorterj_percentage}
\end{figure}

To the best of the author's knowledge, the shorter-J phenomenon is currently the main source of non-standard waveforms except slurs and notches, and it is fatal to the traditional approach of J-location based on significant J peaks. Figure \ref{fig_shorterj_percentage} illustrates the percentage of shorter-j in 40 records in the dataset. Especially in the X1012 record, shorter-J occupies 60\%. 

\section{MATERIALS and METHODS}

The shorter-J and the lost-K phenomena reduce the performance of most algorithms mathematically depend on the prominence of the shape of the I-J-K waveforms. To address this challenge, we propose three signal transform methods. Two of them recover the lost K valleys. The third reduces the shorter J phenomenon.

In the following of this paper, for simplicity, we use the term 'raw' to refer to the unfiltered raw BCG signal, we use 'bcg' to indicate the band-pass filtered BCG signal using a 2-10 Hz range, as described in \cite{Samuel2019Ejection}. It should be noted that the filter band in this case differs from that used in \cite{junnila2005emfi,junnila2006wireless,Akhbardeh2007bseg++} which is 2-20HZ.  However, the 2-10HZ and 2-20HZ filtering exhibit negligible difference in shape, as the majority of the effective BCG signal energy lies within the 2-10 Hz frequency band\cite{Lixin2016JpeakExtraction}. The 'bcj' designation is used to depict the band-pass filtered BCG signal utilizing the narrow J-peak band \cite{Lixin2016JpeakExtraction}, which is 2-6 Hz in this paper.

\subsection{Transform based on curvature}

In examining the BCG waveforms with the lost K-valley, we discovered that the curvature of the signal could be utilized to facilitate the recovery of the lost-K valley. In light of this discovery, we propose two signal transform methods that exploit the curvature attribute of the BCG signal. The first transform (equation \ref{eq_bcc}),named bcc(the letter c comes from the initial letter of the word curvature) uses the curvature formula created by Gauss in 1827 \cite{Gauss1827Disquisitiones}.  The second transformation (equation \ref{eq_bcd}),called bcd(the letter d comes from the first letter of the word derivative) uses the inverted second derivative. The $\alpha $ in both transformations is a constant parameter that scales the transformed data so that the resulting waveform is in the same range as the original waveform.

\begin{center}
\begin{equation}
x_{bcc} = -1 \times \frac{\left |{x_{bcj}}''\right | }{(1+{x_{bcj}}'^{2})^{3/2}} \times \alpha 
\label{eq_bcc}
\end{equation}
\end{center}

\begin{center}
\begin{equation}
x_{bcd} = -1 \times {x_{bcj}}'' \times \alpha 
\label{eq_bcd}
\end{equation}
\end{center}

\begin{figure}[htbp]
\centerline{\includegraphics[width=\linewidth]{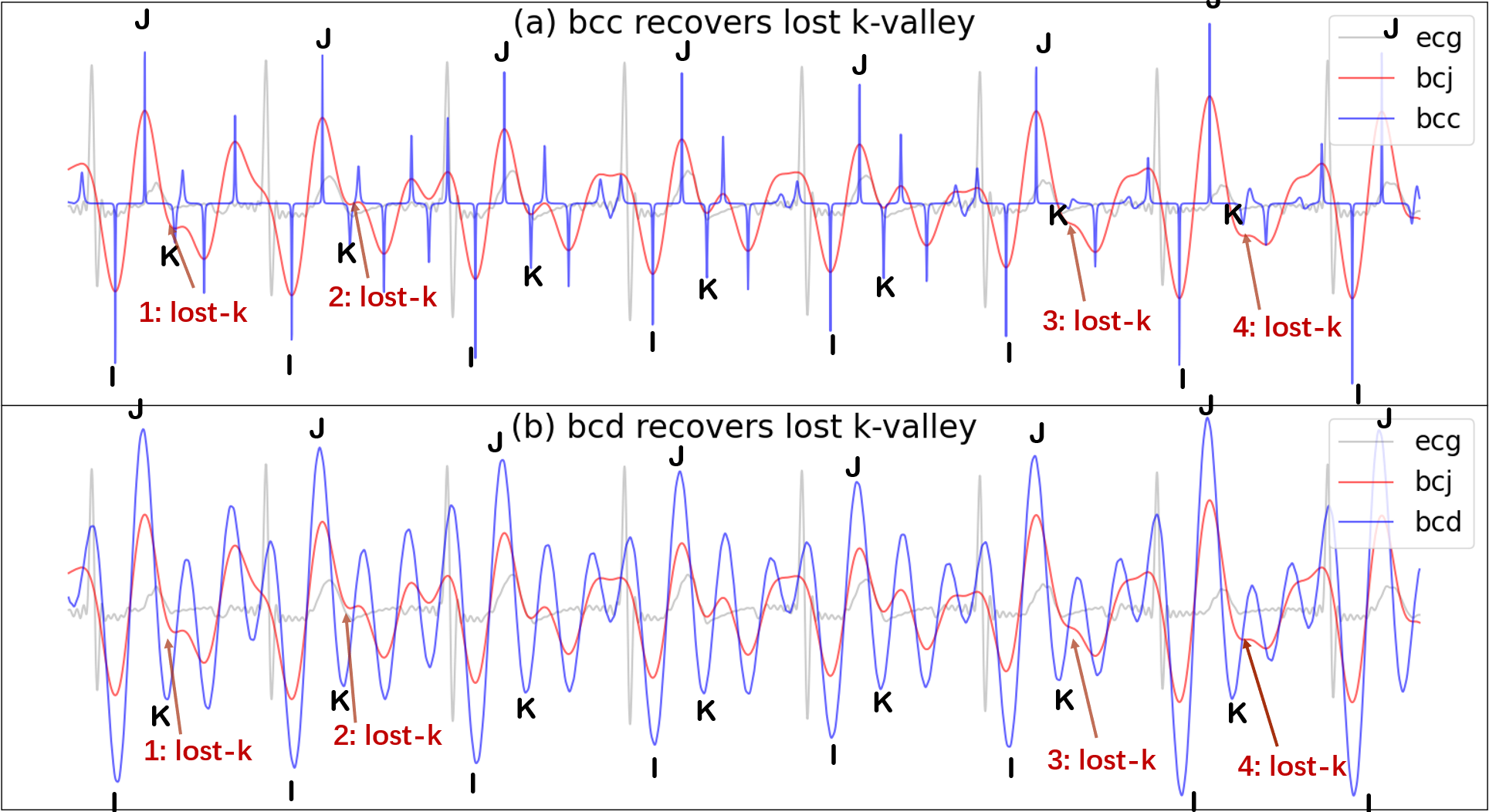}}
\caption{Comparison of  bcj .vs. bcc(a) and bcj .vs. bcd(b)  for a  recording fragment from X1021. Both bcc and bcd show lost-K recoverability. It is notable that bcd exhibits better lost-K recoverability compared to bcc, successfully recovering all lost-K(1,2,3,4), whereas bcc barely recovers lost-K(3,4). }
\label{fig_bcc_bcd_isbetter_lostk}
\end{figure}

The benefit of bcd and bcc transformation is evident. They significantly recover the lost K-Valleys from bcj.  Figure \ref{fig_bcc_bcd_isbetter_lostk} presents a comparison between bcc,bcd and bcj for a dataset fragment. which demonstrates  that bcc,bcd exhibits superior lost-k recoverability. It is important to note that the recovered K-peaks will not be identical to the actual K-peaks. However, the recovered K-peaks can provide a more meaningful J-K delay.

\begin{figure}[htbp]
\centerline{\includegraphics[width=\linewidth]{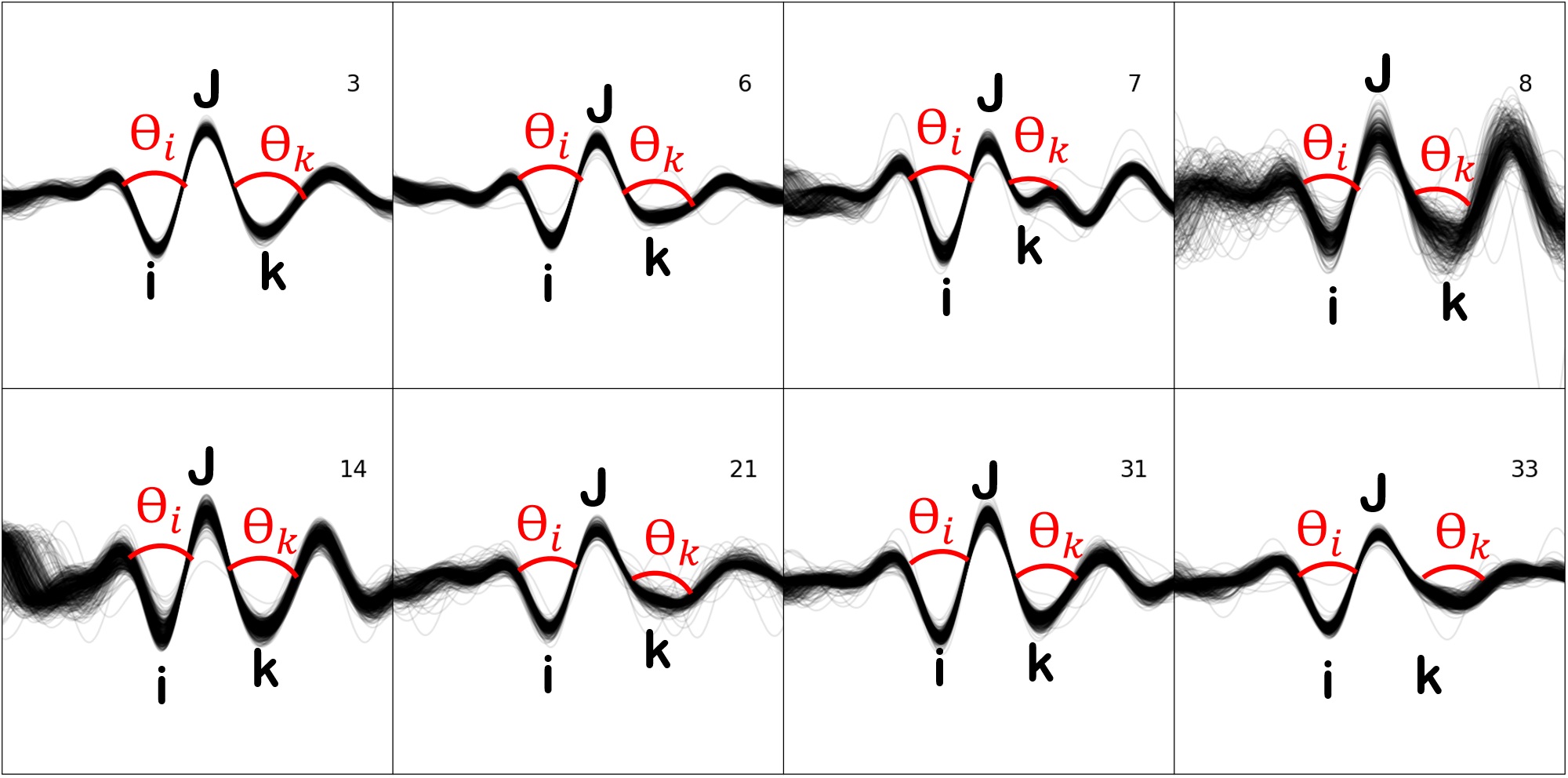}}
\caption{ Stacked view of BCG cycle waveforms from 8 representative recordings. All have a sharper I-Valley($\theta_{i}$ < $\theta_{k}$) due to smaller J-K-L amplitude(X1003,X1031), flattened J-K-L waveforms(X1006,X1021,X1033), insignificant K waveforms(X1007) or larger J-K-L delay(X1008,X1014)}
\label{fig_curvature_why}
\end{figure}

In addition, another interesting phenomenon was observed. In a significant proportion of the bcg waveforms, the H-I-J angle is generally sharper than J-K-L angle as shown in Figure \ref{fig_curvature_why}. We call this phenomenon as sharper-I phenomenon. This phenomenon has been observed in the majority of the records in the dataset. 

\begin{figure}[htbp]
\centerline{\includegraphics[width=\linewidth]{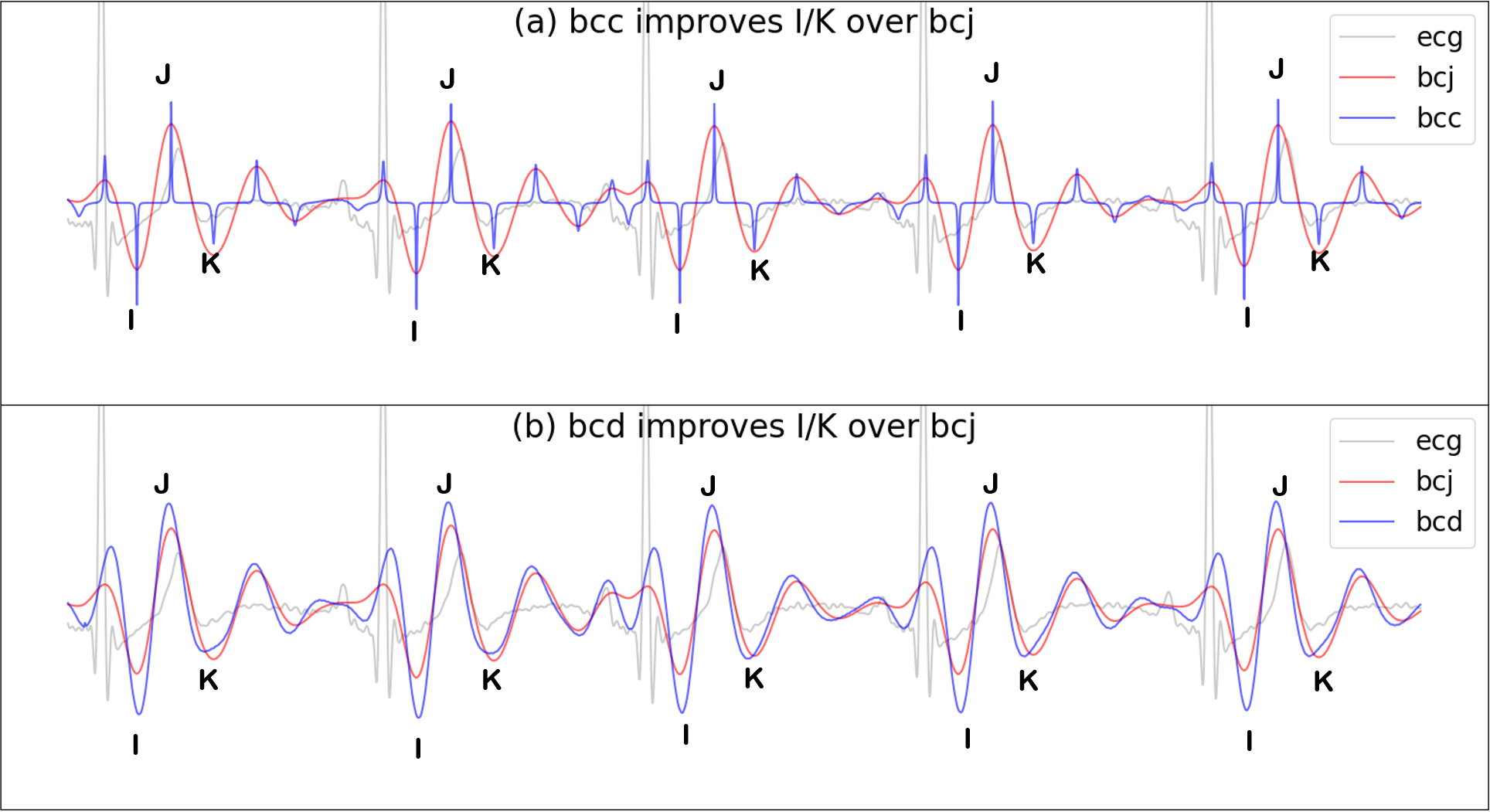}}
\caption{Comparison of  bcj .vs. bcc(a) and bcj .vs. bcd(b)  for a X1003 recording fragment. Both bcc and bcd enhance the I-Valley over the K-Valley, and improve the I-J-K waveform prominence over bcj.}   
\label{fig_bcc_bcd_isbetter_prominence}
\end{figure}

Although research in \cite{Javaid2015Quantifying} highlighted that BCG waveform morphology is distorted by on-bed posture of the subject. The reason behind the phenomenon is beyond the discussion of this paper. The interesting part is that bcc and bcr exploit the sharper-I phenomenon, and enhance the prominence of I-Valley over the k-Valley, significantly improve the I-J waveform prominence. Figure \ref{fig_bcc_bcd_isbetter_prominence} presents a comparison that demonstrates that bcc and bcd exhibit a superior prominence for the I-Valley in comparison to bcj. 

\subsection{Transform based on Coarse Signal}

\begin{center}
\begin{equation}
bcr_{i} = \frac{bcj_{i} \times C_{i}  }{C_{i-T}} \quad
C_{i} = max(BPF(|bcj_{i}^P| + B),1)
\label{eq_bcr}
\end{equation}
\end{center}

\begin{figure}[htbp]
\centerline{\includegraphics[width=\linewidth]{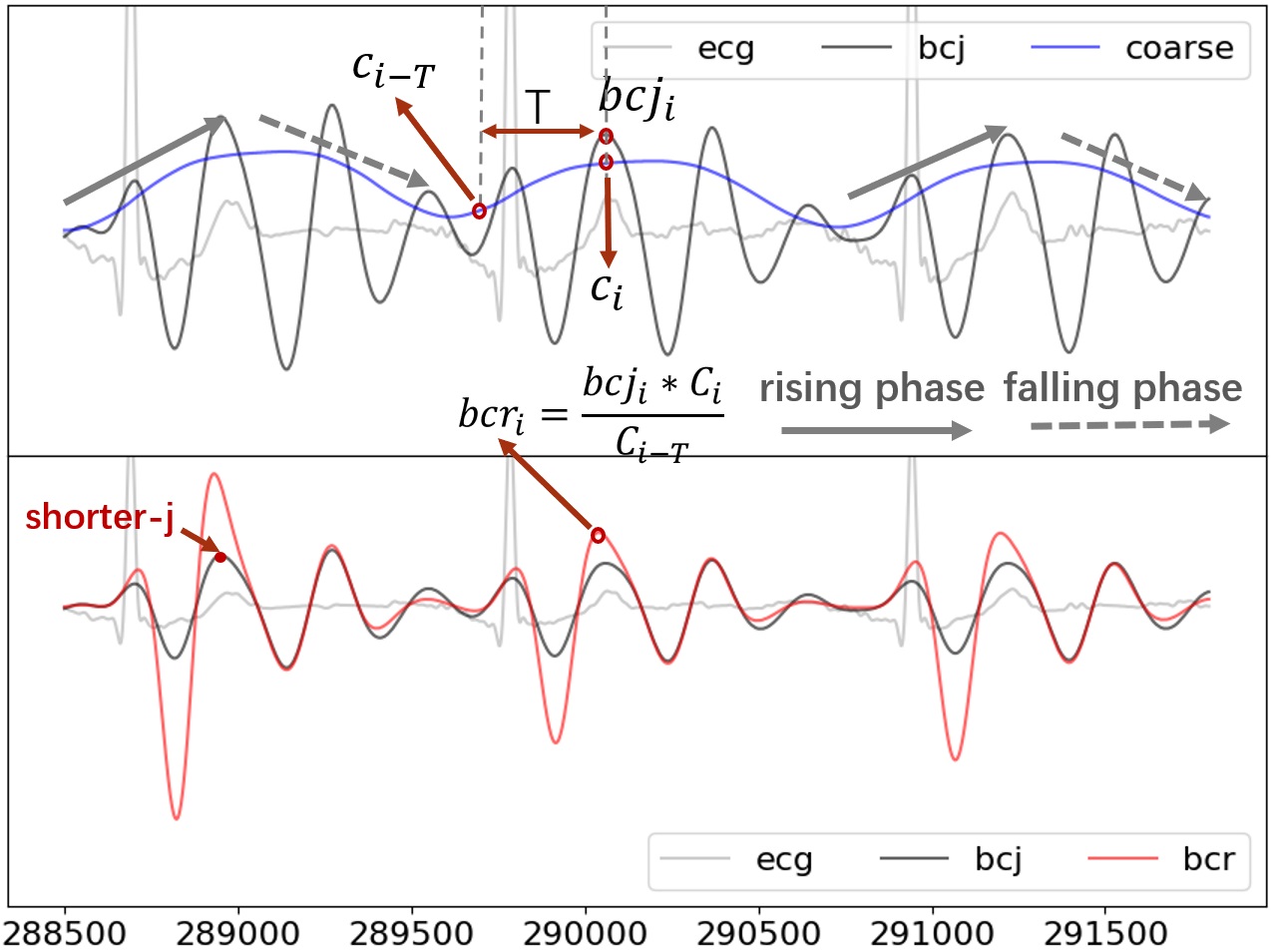}}
\caption{A fragment from participant X1011, (top) bcj and the corresponding coarse signal.(bottom) bcj and the transformed bcr signal. An example shows how bcr is calculated by bcj and the coarse(C) signal. Note that, the shorter-j phenomenon indicated by the arrow is eliminated in bcr. }
\label{fig_whybcr}
\end{figure}

The coarse BCG signal is extracted by a narrow band-pass filter of the absolute value of the bcj signal. the coarse signal is employed in in \cite{Akhbardeh2007bseg++} and \cite{Samuel2019Ejection} as an auxiliary signal  used to determine the position of the J-peak of the BCG signal. The coarse signal is composed of the rising and falling phases, as illustrated in Figure \ref{fig_whybcr}. Observation indicates that the rising phase of the coarse signal signalise the BCG cycle origin, and, the signal in the rising phase should be exaggerated, while the signal in the falling phase should be suppressed. This approach would effectively eliminate the shorter-j phenomenon.

based on the above observation, a new transform method called bcr(the letter r is from the word 'Rising
) is proposed.
Equation \ref{eq_bcr} shows how this transformation is accomplished, where $C_{i}$ denotes the ith sample value of coarse signal. The time window width T (unit as ms) refers to half the Peak-to-Peak interval in coarse signal C, T should be obtained adaptive during runtime, e.g. by rough calculation of IBI from the FFT spectrum.  However, in this paper, T is set to a constant value of 300 in order to facilitate the reproduction of this work by others. 300ms is feasible because HR rate in most dataset recordings is less than 100 Bpm (IBI as 600ms) . 
The P in equation (3) denotes a power applied to the signal, would be optionally be 2 (squared) or 3 (cubed\cite{Smrcka2005NewMethods,Samuel2019Ejection}) to enhance the signal further. in this paper, P is set as 1.  

The symbol B represents a bias parameter that is added to $C_{i}$ because $C_{i}$, as a filtered signal, can be less than or equal to 0. Adding B to $C_{i}$ prevents the transformed signal from being very large. In a sense, adjusting B also adjusts the gain of the transform. In industrial practice, the value of B can be set as an a priori constant or as an adaptive runtime parameter. In this paper, for the sake of simplicity, B is set for each dataset  record to the absolute mean value of the C signal over the entire record.

\begin{figure}[htbp]
\centerline{\includegraphics[width=\linewidth]{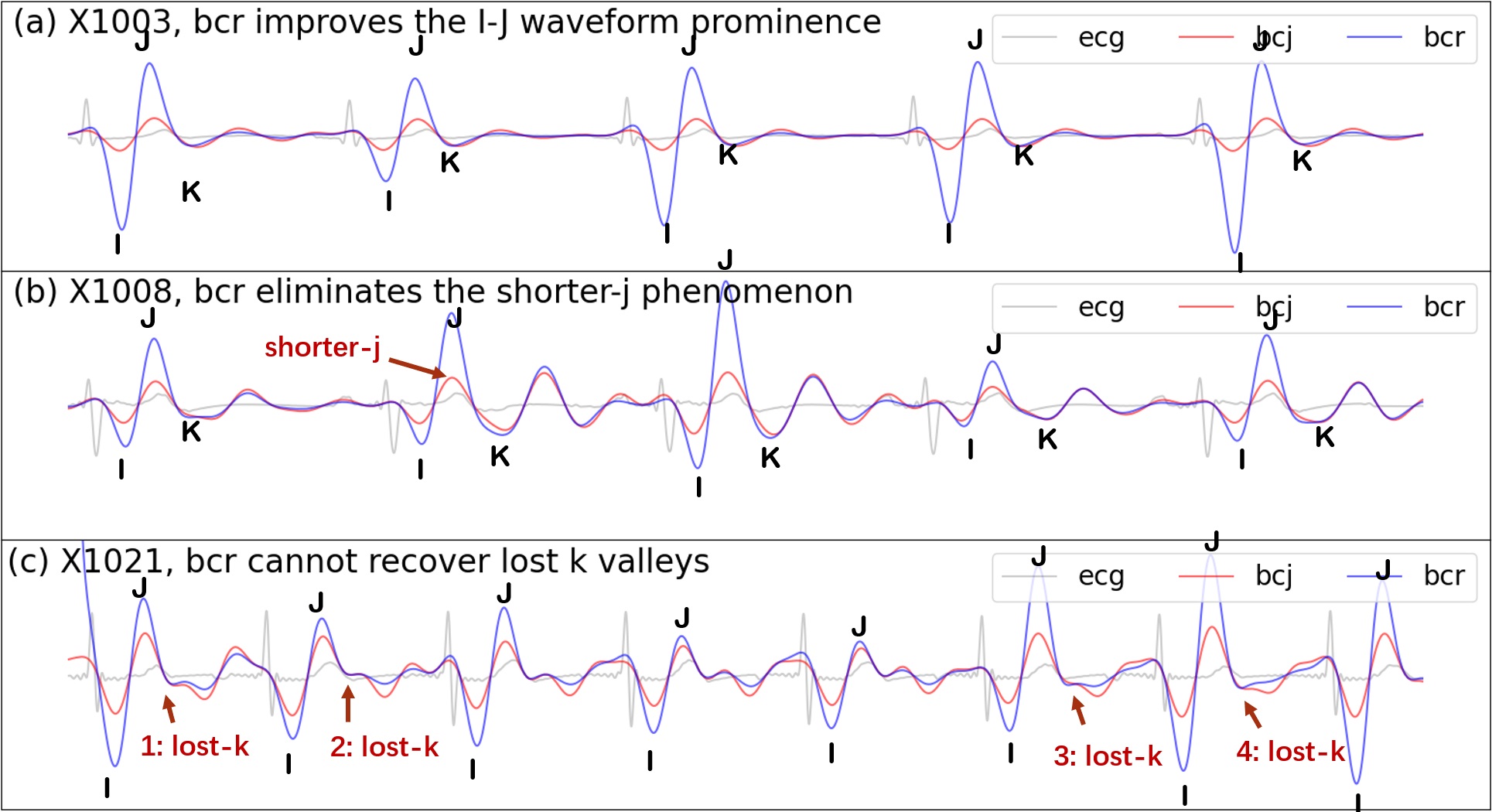}}
\caption{Comparison of bcr .vs. bcj for fragments from X1003(a), X1008(b), X1021(c). The bcr exihibits superior prominence over bcj(a) and eliminates shorter-j phenomenon(b). However, it fails to recover the lost-K(c).}
\label{fig_bcr_isbetter}
\end{figure}

Figure \ref{fig_whybcr} shows how an example of the transform. Figure \ref{fig_bcr_isbetter} compares bcr and bcj over record X1003(a), X1008(b) and X1021(c). The comparison shows that bcr significantly improves the significance of the I-J waveform compared to bcj. and eliminates the specific shorter-j phenomenon, however, it fails to recover the lost k-valleys.

\section{EVALUATION and RESULTS}

This section evaluates the performance of the proposed transform methods in four ways: 1. waveform prominence, 2. precision and recall of the J-peak location,3.  recoverability of J-K delay timing, 4. percentage of shorter-J.  

\subsection{Prominence}

Three metrics are used in this section to assess waveform prominence: i-prominence, j-prominence, ij-prominence. The i-prominence is calculated by dividing the I amplitude by the minimum of the amplitudes of neighboring valleys (e.g. E,G,K,M) within T time around the I-Valley. The j-prominence is calculated by dividing the J amplitude by the maximum amplitude of peaks (e.g. F,H,L,N) within T time around the J peak. The ij prominence is calculated by calculating the i-j amplitude divided by the maximum of all rising waveforms (e.g. E-F,G-H,K-L,M-N) within T time around the I-J waveform. The parameter T is set as 500 ms in this paper. 

\begin{figure}[htbp]
\centerline{\includegraphics[width=\linewidth]{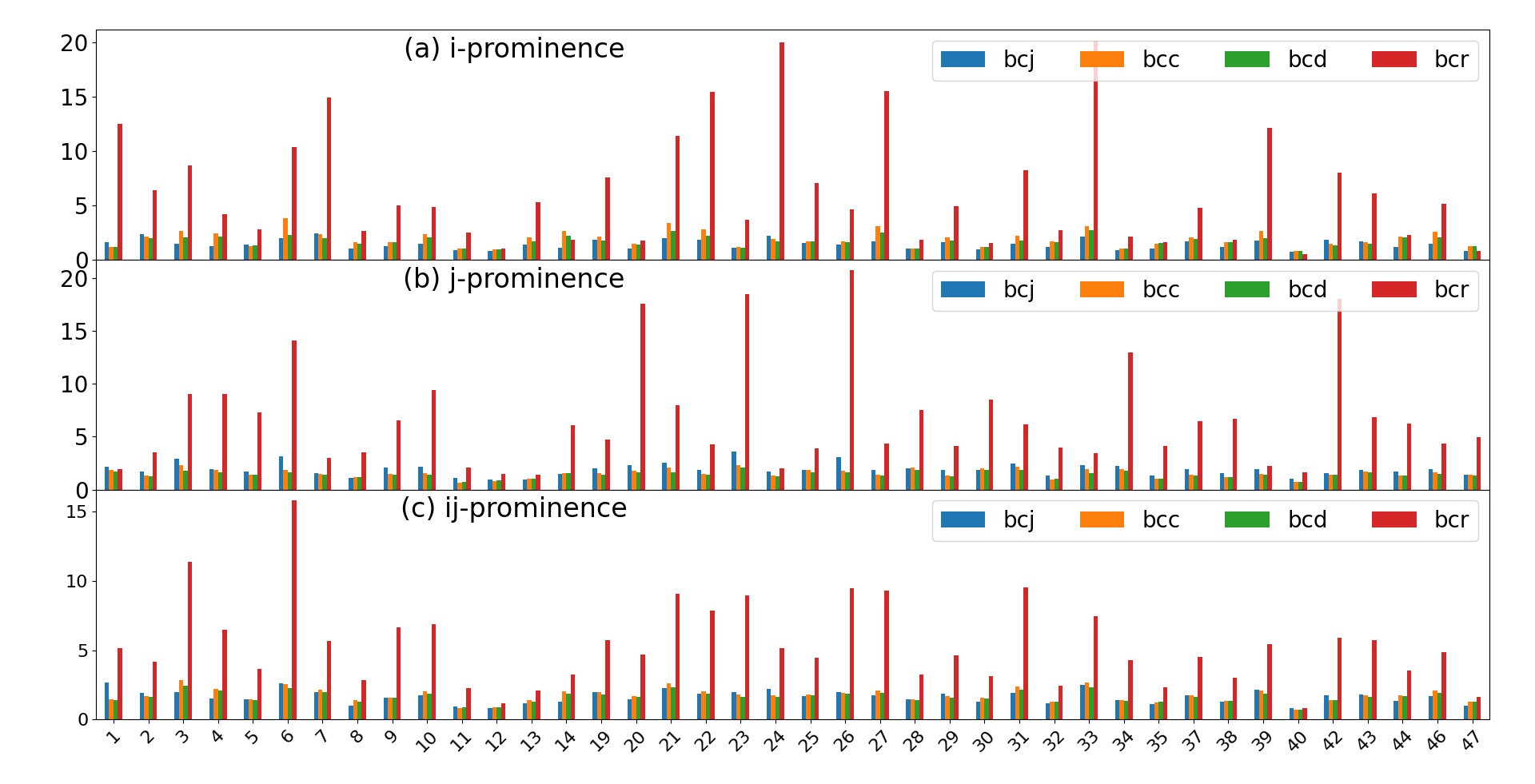}}
\caption{Comparison of i-prominence, j-prominence, ij-prominence between bcj,bcc,bcd and bcr. }
\label{fig_prominence}
\end{figure}

Figure \ref{fig_prominence} shows the results of the evaluation. Compared to bcj, both bcc and bcd show similar performance in the three metrics and do not improve the prominence much. However, most of their i-prominence is better than that of bcj (29 out of 40 records). Notably, All three metrics of bcr are far better than bcj, bcc and bcd in most datasets, except X1012 and 1040. 

\subsection{Recall and Precision}
We compare the recall and precision of the four signals bcj,bcc,bcd,bcr for j-peak locating under simple rules as following, each uses its own best rule : 
 bcj uses max(j): J is identified as waveform peak with max amplitude in neighborhood peaks. 
 bcc,bcd use min(j): I is identified as waveform valley with minimum amplitude in neighborhood valleys.
 bcr uses max(ij): J is identified as waveform valley-to-peak with max i-j amplitude in neighborhood upward waveforms(e.g. e-f,g-h,k-l,m-n).

\begin{figure}[htbp]
\centerline{\includegraphics[width=\linewidth]{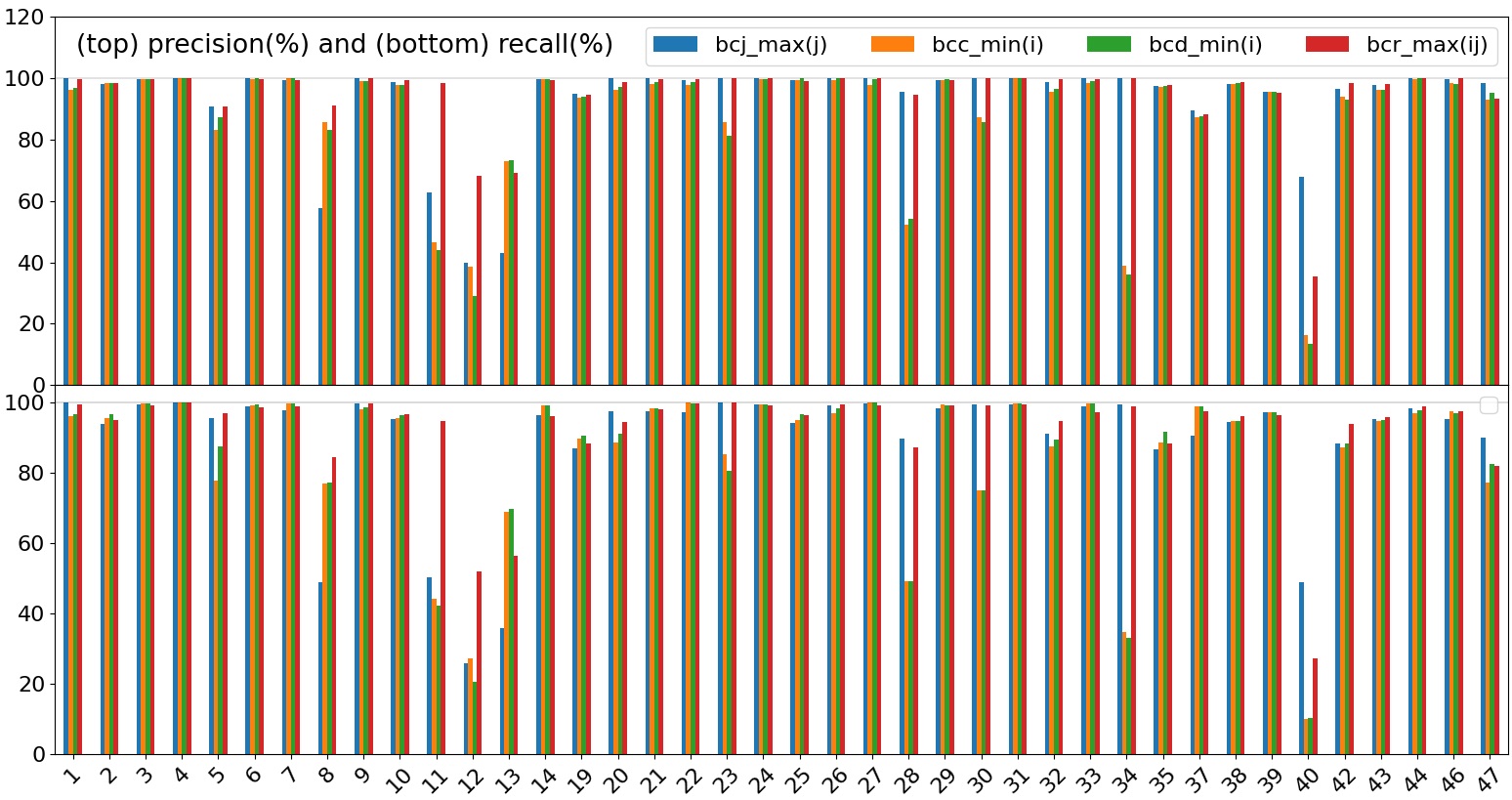}}
\caption{Comparison of recall and precision between bcj,bcc,bcd and bcr with simple rules,respectively.}
\label{fig_recall_and_precision}
\end{figure}

The performance of transform methods are evaluated by the precision, recall, as shown below:
\begin{center}
\begin{equation}
Recall = \frac{TP }{TP+FN}, Precision = \frac{TP }{TP+FN}
\end{equation}
\end{center}
where TP represents the ground truth of J-peaks, validated by the time-alighned ECG signal, FP is the J-peaks detected by the rules but failed to be the ground truth j-peak, FN is the J-peaks failed to be detected by the rules.As illustrated in figure (\ref{fig_recall_and_precision}), the bcr$\underline{~}$max(ij) method shows the best performance than bcc$\underline{~}$min(i),bcd$\underline{~}$min(i),and bcj$\underline{~}$max(j) for all records except X1040,which has a larger J-K amplitude than I-J amplitude in waveform shape. 

Additionally , it is seen that the performance of bcc,bcd depends on the feature of sharper-I in waveform shape , which does not apply to all BCG waveform shapes. For most records, bcc$\underline{~}$min(i),bcd$\underline{~}$min(i) have about the same performance with bcj$\underline{~}$max(j). The bcc$\underline{~}$min(i),bcd$\underline{~}$min(i) show some improvement for X1008,X1013, however, the effect slips for X1011,X1012 and  significant decrease for X1028,X1034. The reason is that X1008,X1013 have sharp-I curvature, while X1011,X1012,X1028,X1034 do not (\ref{fig:waves_stack}).

\subsection{K-valley recovery}

\begin{figure}[htbp]
\centerline{\includegraphics[width=\linewidth]{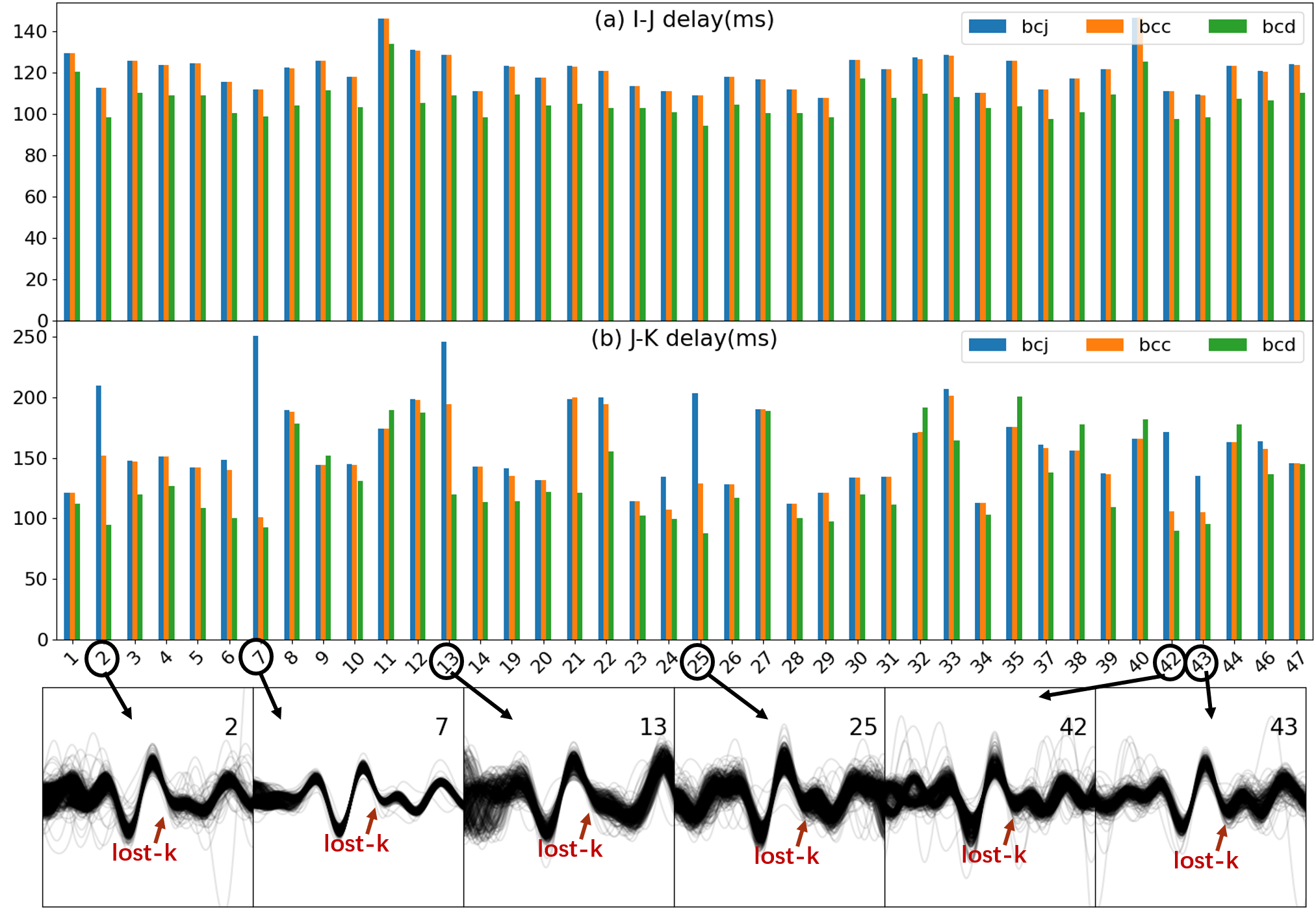}}
\caption{ Comparison of I-J delay(a) and J-K delay(b) between bcj,bcc,and bcr.(bottom)Stackview of BCG cycle waveforms for some obvious K-valley recover records. Arrows show the locations where lost-k phenomenon would easily occur.}
\label{fig_ijt_and_jkt}
\end{figure}

Figure (\ref{fig_ijt_and_jkt}) shows the IJ delay and the JK delay in 40 records for bcj, bcc, and bcd.  There is little difference in the IJ delay between bcc,bcd, and bcj. In contrast, bcc and bcd show a large improvement in JK delay over bcj for several records, notably X1002, X1007, X1025, X1042, and X1043. In these records, the JK delay is reduced from more than 200ms to less than 150ms. Although there is no ground truth for the JK delay, the improvement can be validated by the JK delay predicted by the model in \cite{kim2016ballistocardiogram}.

\subsection{shorter-j elimination}

\begin{figure}[htbp]
\centerline{\includegraphics[width=\linewidth]{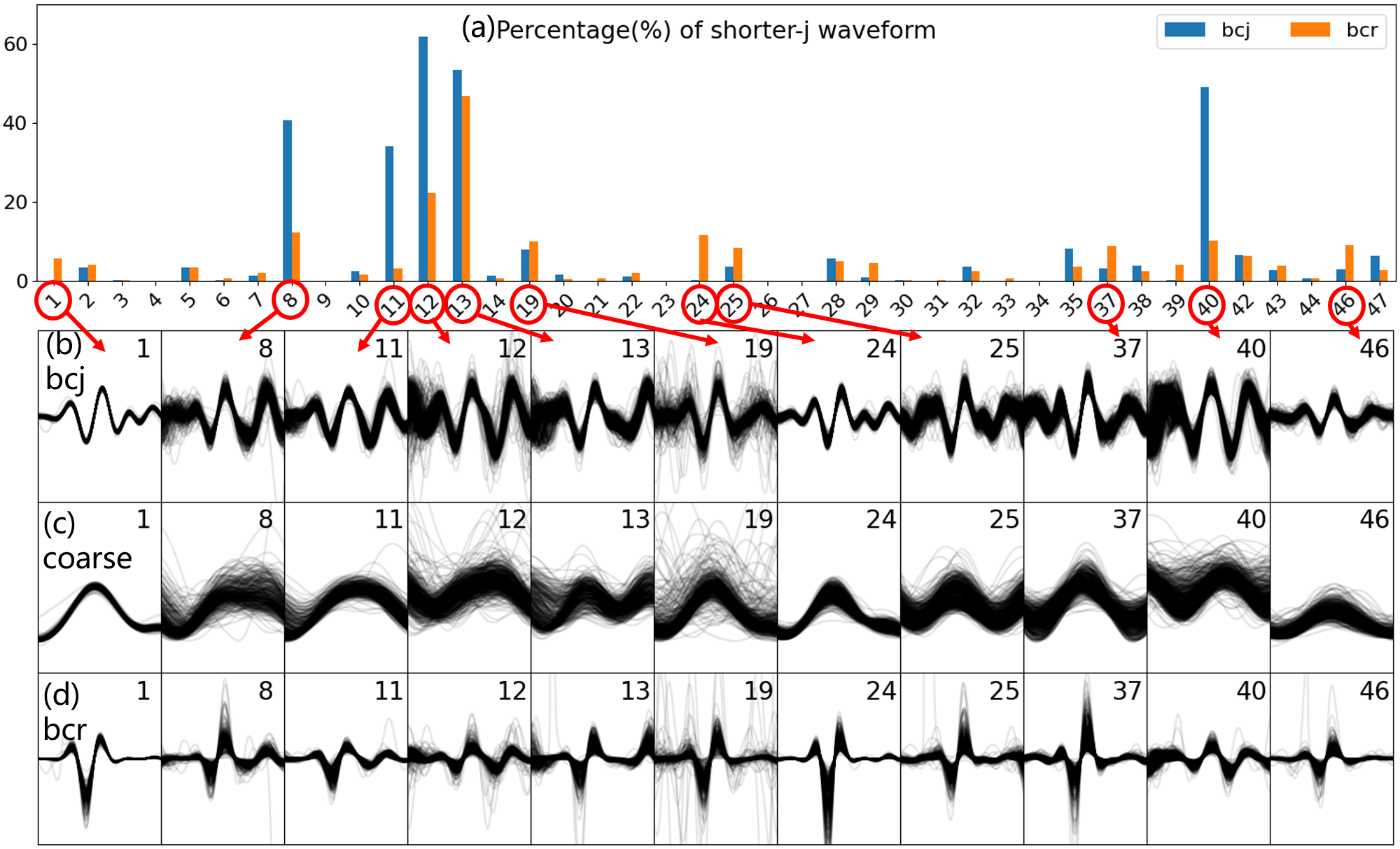}}
\caption{(a) Shorter J elimination of bcr over bcj. (b) Stack view of BCG cycles for some selected cases from bcj, (c) the corresponding stack view of the coarse signal of bcj, and (d) the corresponding stack view of bcr.}
\label{fig_shorterjratio_improvement}
\end{figure}

Figure (\ref{fig_shorterjratio_improvement}) shows the elimination of the shorter-J phenomenon of bcr compared to bcj. For dataset records X1008, X1011, and X1040, bcr performs well, significantly reducing the proportion of shorter-J phenomenon. For X1012, bcr does not perform well because the rising phase of the corresponding coarse signal is delayed, so the latter part of the BCG cycle is amplified to produce a larger L peak. For X1001,X1024,X1025,X1037,X1046, the bcr transform elevates the amplitude of the H-wave peaks, resulting in some H peaks being larger than  J peaks, which worsens the shorter-J issue. However, this degradation  does not affect BCG cycle segmentation methods using the more prominent I-peaks rather than J-peaks.

\section{DISCUSSION}

It is important to note that the I, J, K positions calculated from bcc, bcd, bcr is only a rough estimate of the real position. In practice, it is necessary to find the corresponding more accurate I,J,K position from bcj (or bcg) in the corresponding neighborhood.  The recovered K-valleys by bcc and bcd provide a more meaningful J-K delay, but further research is needed to validate the recovered K valley position. The bcr transform may yield less accurate results than bcj in some cases, but it significantly improves the prominence of the I-J-K complex. This improvement is particularly effective when using machine learning or deep learning methods. Furthermore, the more prominent I-valley can be used as an alternative to the J-peak to segment the BCG waveform.

\section{CONCLUSION}

This paper  describes two phenomena in non-standard BCG signals: the shorter-J and the lost-K phenomenon.These two phenomena reduce the prominence of I-J-K waveform.  Three signal transform methods are proposed to  address this challenge. Evaluation results show that the bcc and bcd transform can recover the lost-K valleys, improving the J-K delay, and the bcr transform significantly reduces the shorter-J phenomenon and  enhances the I-J prominence , improving the performance of J-peak based locating methods, especially for non-standard BCG data.

\vspace{12pt}
\tiny
\printbibliography

\end{document}